\documentclass[12pt,cite,epsf,epsfig]{article}
\usepackage{epsfig}

\begin{document}

\author{S. Dev\thanks{dev5703@yahoo.com}, Shivani Gupta\thanks{shiroberts\_1980@yahoo.co.in} and Radha Raman Gautam\thanks{gautamrrg@gmail.com}}

\title{Parallel hybrid textures of lepton mass matrices}
\date{\textit{Department of Physics, Himachal Pradesh University, Shimla 171005, India.}\\
\smallskip}

\maketitle
\begin{abstract}
We analyse the parallel hybrid texture structures in the charged lepton and the neutrino sector. These parallel hybrid texture structures have physical implications as they cannot be obtained from arbitrary lepton mass matrices through weak basis transformations. The total sixty parallel hybrid texture structures can be grouped into twelve classes, and all the hybrid textures in the same class have identical physical implications. We examine all the twelve classes under the assumption of non-factorizable phases in the neutrino mass matrix. Five out of the total twelve classes are found to be phenomenologically disallowed. We study the phenomenological implications of the allowed classes for 1-3 mixing angle, Majorana and Dirac-type $CP$ violating phases. Interesting constraints on effective Majorana mass are obtained for all the allowed classes. 
\end{abstract}

\section{Introduction}
The origin of fermion masses and mixing apart from $CP$ violation remains one of the least understood aspects of the Standard Model (SM) of fundamental particle interactions. In the SM, fermion masses and mixing angles are free parameters and span a wide range of values and to accommodate such a diverse mass spectrum, the Yukawa couplings must span over five orders of magnitude. In the SM, neutrinos are massless but experiments performed during the last two decades have established that neutrinos have small but nonvanishing masses which leads to a further increase in the number of free parameters. Thus, the main theoretical challenges are to reduce the number of free parameters in the Yukawa sector and to obtain tiny neutrino masses but large mixing angles. The proposals aimed at reducing the number of free parameters and, thereby, restricting the form of the mass matrices include the presence of texture zeros \cite{1, 2, 3, 4, 5, 6}, requirement of zero determinant \cite{7} and the zero trace condition \cite{8}. In addition, the presence of vanishing minors \cite{9} and the simultaneous existence of a texture zero and a vanishing minor have recently \cite{10} been investigated. \\
Attempts have been made to understand the observed pattern of quark/lepton masses and mixings by introducing flavor symmetries (Abelian as well as non-Abelian) which naturally leads to such texture structures. To be more specific, texture zeros and flavor symmetries have yielded quantitative relationships between fermion mass ratios and flavor mixing angles. A unified description of flavor physics and $CP$ violation in the quark/lepton sectors can be achieved by constructing a low energy effective theory with the SM gauge symmetry and some discrete non-Abelian family symmetry and, subsequently, embedding this theory into Grand Unified Theory (GUTS) models like SO(10) \cite{11}. The search for an adequate discrete symmetry has mainly focussed on the minimal subgroups of SO(3) or SU(3) with at least one singlet and one doublet irreducible representation to accommodate the fermions belonging to each generation. One such subgroup, for example, is the quaternion group $Q_8$ \cite{12} which not only accommdates the three generations of fermions but also explains the rather large difference between the values of the 2-3 mixing in the quark and lepton sectors. Quaternion symmetry like some other discrete symmetries leads to nontrivial relationships amongst the nonvanishing elements of the mass matrix. Such textures with equalities between different elements alongwith some vanishing elements have been referred to as hybrid textures. Detailed phenomenological analyses of hybrid textures with one texture zero and an equality in the neutrino mass matrix in the flavor basis have been reported \cite{13} earlier. However, the investigation of the neutrino mass matrix in the diagonal charged lepton basis can only be regarded as a precursor of a more general study where both the charged lepton and the neutrino mass matrices are non-diagonal. In the present work, we examine the implications of parallel hybrid texture structures of both the charged lepton and the neutrino mass matrices in a nondiagonal basis assuming the charged lepton mass matrices to be hermitian. However, it is pertinent to emphasize here that hermitian charged lepton mass matrices cannot be obtained within the framework of the standard electroweak gauge group. Furthermore, parallel hybrid texture structures for lepton mass matrices can only be ensured by imposing discrete non-Abelian lepton flavor symmetries.  There exist a rather large, sixty to be precise [Table 1], number of possible hybrid textures with one texture zero and one equality between mass matrix elements. However, we find that these sixty hybrid texture structures can be grouped into twelve classes such that all the hybrid textures belonging to a particular class have the same physical implications. We examine the phenomenological implications of all the twelve classes of neutrino mass matrices with hybrid texture structure under the assumption of non-factorizable phases in the neutrino mass matrix.  \\
 The $CP$ violation in neutrino oscillation experiments can be described
through a rephasing invariant quantity, $J_{CP}$ \cite{14} with
$J_{CP}=Im(U_{e1}U_{\mu2}U_{e2}^*U_{\mu1}^*)$. In the
parameterization adopted here, $J_{CP}$ is given by
\begin{equation}
J_{CP} = s_{12}s_{23}s_{13}c_{12}c_{23}c_{13}^2 \sin \delta.
\end{equation}
The effective Majorana mass of the electron neutrino $M_{ee}$ which
 determines the rate of neutrinoless double beta decay is given by
\begin{equation}
M_{ee}= |m_1c_{12}^2c_{13}^2+ m_2s_{12}^2c_{13}^2 e^{2i\alpha}+ m_3s_{13}^2e^{2i\beta}|. 
\end{equation}
This important parameter will help decide the nature of neutrinos. The analysis of $M_{ee}$ will be significant as many neutrinoless
double beta decay experiments will constrain this
parameter. A stringent constraint $|M_{ee}|< 0.35$eV was obtained by the $^{76}Ge$ Heidelberg-Moscow experiment \cite{15} .
 There are large number of projects such as SuperNEMO\cite{16},
 CUORE\cite{17}, CUORICINO\cite{17} and  GERDA\cite{18} which
 aim to achieve a sensitivity below 0.01eV to $M_{ee}$.
 Forthcoming experiment SuperNEMO, in particular, will explore
 $M_{ee}$ $<$ 0.05eV\cite{19}.
The experimental constraints on the neutrino parameters at 1, 2 and
3$\sigma$ \cite{20} are given below:
\begin{eqnarray}
\Delta m_{12}^{2}
&=&7.67_{(-0.19,-0.36,-0.53)}^{(+0.16,+0.34,+0.52)}\times
10^{-5}eV^{2}, \nonumber \\ \Delta m_{23}^{2} &=&\pm
2.39_{(-0.8,-0.20,-0.33)}^{(+0.11,+0.27,+0.47)}\times 10^{-3}eV^{2},  \nonumber \\
\theta_{12}& =&33.96_{(-1.12,-2.13,-3.10)}^{o(+1.16,+2.43,+3.80)},
\nonumber \\ \theta_{23}
&=&43.05_{(-3.35,-5.82,-7.93)}^{o(+4.18,+7.83,+10.32)}, \nonumber \\
\theta_{13} &<& 12.38^{o}(3\sigma).
\end{eqnarray}
The upper bound on $\theta_{13}$ is given by the CHOOZ experiment.
\section{Weak Basis Transformations}
Assuming neutrinos to be of Majorana nature, the most general weak basis transformation (under which lepton mass matrices change but which leaves the gauge currents invariant) is
\begin{equation}
M_l\longrightarrow M_l'=W^{\dagger}M_l W',\  M_{\nu}\longrightarrow M_{\nu}'=W^{T}M_{\nu}W  
\end{equation}
where $W$ and $W'$ are 3$\times$ 3 unitary matrices and $M_l$, $M_{\nu}$ are the charged lepton and the neutrino mass matrices, respectively. 
\subsection{Parallel hybrid texture structures.}
In this section we investigate the possibility of obtaining parallel hybrid texture structures starting from an arbitrary hermitian charged lepton and complex symmetric neutrino mass matrix. We follow the line of argument advanced by Branco \textit{et al.} \cite{21} for parallel four texture zero Ans\"{a}tze. For illustration we choose a specific hybrid texture structure (IA) [Table 1] with a zero at (1,3) position and equality of (1,1) and (2,2) elements:
\begin{equation}
M_l=\left(%
\begin{array}{ccc}
  a_l & b_l & 0 \\
  b_l^* & a_l & e_l \\
 0 & e_l^* & f_l \\
\end{array}%
\right),
 M_\nu=\left(%
\begin{array}{ccc}
  a_\nu & b_\nu & 0 \\
  b_\nu& a_\nu &e_\nu \\
  0 &e_\nu  & f_\nu \\
\end{array}
\right)
\end{equation}
where $M_l$ is hermitian and $M_{\nu}$ is complex symmetric.
We can rephase $M_l$ and $M_{\nu}$ such that
\begin{equation}
M_l\longrightarrow X^{\dagger}M_l X,\  M_{\nu}\longrightarrow X^{T}M_{\nu} X  
\end{equation}
where $X\equiv$ diag($e^{i\phi_1}$,$e^{i\phi_2}$,$e^{i\phi_3}$) and we can choose $\phi_i$ such that $M_l$ becomes real. We can use the remaining freedom to remove one phase from $M_{\nu}$.
It is instructive to enumerate the number of free parameters in the above two parallel hybrid texture structures. The charged lepton mass matrix after rephasing is left with four real parameters. There are seven free parameters in the neutrino mass matrix (four real parameters and three phases). In total, we have eleven free parameters in $M_l$ and $M_{\nu}$ whereas in the leptonic sector considering neutrinos to be Majorana particles there are twelve physical parameters (six lepton masses, three mixing angles and three phases) for three generations of neutrinos. Thus, starting from an arbitrary hermitian $M_l$ and complex symmetric $M_\nu$, we cannot obtain parallel hybrid texture structures through weak basis transformations as the number of free parameters for such texture structures is less than twelve. This implies that these parallel texture structures have physical implications. However, if the condition of hermiticity in $M_l$ is removed then these parallel texture structures can be obtained through weak basis transformations and will, thus, have no physical implications as the total number of free parameters now in $M_l$ and $M_{\nu}$ are greater than twelve. In our analysis, we consider $M_l$ to be hermitian. 
 
\subsection{Weak basis equivalent classes of hybrid textures.}
Different parallel hybrid texture structures of the charged lepton and neutrino mass matrices can be related by a weak basis transformation. The implications of these hybrid textures which are related by such a transformation are exactly the same. This weak basis transformation can be performed by a permutation matrix P as  
\begin{eqnarray}
M'_l=P^TM_lP, \nonumber \\
M'_\nu = P^TM_\nu P.
\end{eqnarray}
which changes the position of the texture zero and an equality but preserves the parallel structure of charged lepton and neutrino mass matrices. The permutation matrix P belongs to the group of six permutation matrices. We find that all sixty hybrid textures [Table 1] fall into twelve distinct classes when operated by these permutation matrices. The different classes are shown in Table 1. However, this type of classification is not possible in the flavor basis \cite{13}, since, such a weak basis transformation will render $M_l$ non diagonal.

\begin{table}[b]
\begin{footnotesize}
\begin{center}
\begin{tabular}{|c|c|c|c|c|c|c|}
 \hline
  Class & A & B  & C & D & E & F \\
 \hline
 I  & $\left(
\begin{array}{ccc}
a & b & 0 \\  &a & e \\ &  & f
\end{array}
\right)$& $\left(
\begin{array}{ccc}
a & b & c \\  & a & 0 \\ &  & f
\end{array}
\right)$  &  $\left(
\begin{array}{ccc}
a & 0 & c \\  & d & e \\ &  & a
\end{array}
\right)$ & $\left(
\begin{array}{ccc}
a & b & c \\  & d & 0 \\ &  & a
\end{array}
\right)$ & $\left(
\begin{array}{ccc}
a & 0 & c \\  & d & e\\ &  & d
\end{array}
\right)$&$\left(
\begin{array}{ccc}
a & b & 0 \\  & d & e\\ &  & d
\end{array}
\right)$ \\
\hline
II & $\left(
\begin{array}{ccc}
0 & b & c \\  & d & e \\ &  & d
\end{array}
\right)$ & $\left(
\begin{array}{ccc}
a & b & c \\  & 0 & e \\ &  & a
\end{array}
\right)$&$\left(
\begin{array}{ccc}
a & b & c \\  &a & e \\ &  & 0
\end{array}
\right)$& - &-&-\\
\hline
III  & $\left(
\begin{array}{ccc}
0 & b & b \\  & d & e \\ &  & f
\end{array}
\right)$ & $\left(
\begin{array}{ccc}
a & b & c\\  & 0 & c \\ &  & f
\end{array}
\right)$ & $\left(
\begin{array}{ccc}
a & b & c \\  & d & c \\ &  & 0
\end{array}
\right)$& - &- &- \\
\hline
IV & $\left(
\begin{array}{ccc}
a & a & c \\  &0 & e \\ &  & f
\end{array}
\right)$ & $\left(
\begin{array}{ccc}
a & b & a \\  &d & e \\ &  & 0
\end{array}
\right)$ &  $\left(
\begin{array}{ccc}
0 & b & c\\  &b & e \\ &  & f
\end{array}
\right)$  & $\left(
\begin{array}{ccc}
a & b &c \\  &d & d \\ &  & 0
\end{array}
\right)$& $\left(
\begin{array}{ccc}
0 & b & c \\  &d & e \\ &  & c
\end{array}
\right)$ & $\left(
\begin{array}{ccc}
a & b & c \\  &0 & e\\ &  & e
\end{array}
\right)$       \\
\hline
V & $\left(
\begin{array}{ccc}
a & a & c \\  &d & e \\ &  & 0
\end{array}
\right)$ & $\left(
\begin{array}{ccc}
a & b & a \\  &0 & e \\ &  & f
\end{array}
\right)$ &  $\left(
\begin{array}{ccc}
a & b & c\\  &b & e \\ &  & 0
\end{array}
\right)$  & $\left(
\begin{array}{ccc}
0 & b &c \\  &d & d \\ &  & f
\end{array}
\right)$& $\left(
\begin{array}{ccc}
a & b & c \\  &0 & e \\ &  & f
\end{array}
\right)$ & $\left(
\begin{array}{ccc}
0 & b & c \\  &d & e\\ &  & e
\end{array}
\right)$       \\
\hline
VI & $\left(
\begin{array}{ccc}
a & b & c \\  &0& a \\ &  & c
\end{array}
\right)$ & $\left(
\begin{array}{ccc}
a & b & c \\  &d & a \\ &  & 0
\end{array}
\right)$ &  $\left(
\begin{array}{ccc}
0 & b & c\\  &c & e \\ &  & f
\end{array}
\right)$  & $\left(
\begin{array}{ccc}
a & b &c \\  &c& e \\ &  &0
\end{array}
\right)$& $\left(
\begin{array}{ccc}
0 & b & c \\  &d & e \\ &  & b
\end{array}
\right)$ & $\left(
\begin{array}{ccc}
a & b & c \\  &c & e\\ &  & 0
\end{array}
\right)$       \\
\hline
VII & $\left(
\begin{array}{ccc}
a & a &0 \\  &d & e \\ &  & f
\end{array}
\right)$ & $\left(
\begin{array}{ccc}
a &0 & a \\  &d & e \\ &  & f
\end{array}
\right)$ &  $\left(
\begin{array}{ccc}
a & b & c\\  &b & 0 \\ &  & f
\end{array}
\right)$  & $\left(
\begin{array}{ccc}
a & 0 &c \\  &d & d \\ &  & f
\end{array}
\right)$& $\left(
\begin{array}{ccc}
a & b & c \\  &d & 0 \\ &  & c
\end{array}
\right)$ & $\left(
\begin{array}{ccc}
a & b & 0 \\  &d & e\\ &  & e
\end{array}
\right)$       \\
\hline
VIII&  $\left(
\begin{array}{ccc}
a & b & c \\  & d & 0 \\ &  & d
\end{array}
\right)$&$\left(
\begin{array}{ccc}
a & b & 0 \\  & d & e \\ &  & a
\end{array}
\right)$  & $\left(
\begin{array}{ccc}
a & 0 & c \\  & a & e \\ &  & f
\end{array}
\right)$  & - & -&- \\
 \hline
IX  &$\left(
\begin{array}{ccc}
a & b & b \\  & d & 0 \\ &  & f
\end{array}
\right)$ & $\left(
\begin{array}{ccc}
a & b & 0 \\  & d &b \\ &  & f
\end{array}
\right)$  & $\left(
\begin{array}{ccc}
a & 0 & c\\  & d & c \\ &  & f
\end{array}
\right)$ & -&-&- \\
\hline
X & $\left(
\begin{array}{ccc}
a & b & b \\  &0 & e \\ &  & f
\end{array}
\right)$ & $\left(
\begin{array}{ccc}
a & b & b \\  &d & e \\ &  & 0
\end{array}
\right)$ &  $\left(
\begin{array}{ccc}
0 & b & c\\  &d & b \\ &  & f
\end{array}
\right)$  & $\left(
\begin{array}{ccc}
a & b &c \\  &d & b \\ &  & f
\end{array}
\right)$& $\left(
\begin{array}{ccc}
0 & b & c \\  &d & c \\ &  & f
\end{array}
\right)$ & $\left(
\begin{array}{ccc}
a & b & c \\  &0 & c\\ &  & f
\end{array}
\right)$       \\
\hline
XI & $\left(
\begin{array}{ccc}
a & a & c \\  &d & 0 \\ &  & f
\end{array}
\right)$ & $\left(
\begin{array}{ccc}
a & b & a \\  &d & 0 \\ &  & f
\end{array}
\right)$ &  $\left(
\begin{array}{ccc}
a & b & 0\\  &b & e \\ &  & f
\end{array}
\right)$  & $\left(
\begin{array}{ccc}
a & b &0 \\  &d & d \\ &  & f
\end{array}
\right)$& $\left(
\begin{array}{ccc}
a & 0 & c \\  &d & e \\ &  & c
\end{array}
\right)$ & $\left(
\begin{array}{ccc}
a & 0 & c \\  &d & e\\ &  & e
\end{array}
\right)$       \\
\hline
XII & $\left(
\begin{array}{ccc}
a & 0 & c \\  &d & a \\ &  & f
\end{array}
\right)$ & $\left(
\begin{array}{ccc}
a & b & 0 \\  &d & a \\ &  & f
\end{array}
\right)$ &  $\left(
\begin{array}{ccc}
a & 0 & c\\  &c & e \\ &  & f
\end{array}
\right)$  & $\left(
\begin{array}{ccc}
a & b &c \\  &c & 0 \\ &  & f
\end{array}
\right)$& $\left(
\begin{array}{ccc}
a & b & 0 \\  &d & e \\ &  & b
\end{array}
\right)$ & $\left(
\begin{array}{ccc}
a & b & c \\  &d & 0\\ &  & b
\end{array}
\right)$       \\
\hline
\end{tabular}
\caption{All possible parallel hybrid texture structures. The hybrid textures in a class have the same physical implications.}
\end{center}
\end{footnotesize}
\end{table}
\section{Comprehensive analysis of different classes of hybrid textures.}
\subsection{Class I}
All the information regarding lepton masses and mixings is encoded in the hermitian charged lepton mass matrix $M_l$ and the complex symmetric neutrino mass matrix $M_\nu$. First, we analyse the hybrid texture structure in which $M_l$ and $M_\nu$ have the parallel structure with equal (1,1) and (2,2) elements and a texture zero at (1,3) place (Case IA). We study this hybrid texture under the assumption of non-factorizable phases in the neutrino mass matrix $M_\nu$, as it is not always possible to factorize all the phases present in a general complex symmetric mass matrix without unnatural fine tuning of phases \cite{22}. Therefore, $M_l$ and $M_\nu$ are given by 
\begin{equation}
M_l=\left(%
\begin{array}{ccc}
  a_l & b_l & 0 \\
  b_l^* & a_l & e_l \\
 0 & e_l^* & f_l \\
\end{array}%
\right),
 M_\nu=\left(%
\begin{array}{ccc}
  a_\nu & b_\nu & 0 \\
  b_\nu& a_\nu &e_\nu \\
  0 &e_\nu  & f_\nu \\
\end{array}
\right)
\end{equation}
respectively. The hermiticity of $M_l$ requires its diagonal elements $a_l$ and $f_l$ to be real whereas the non diagonal elements $b_l$ and $e_l$ are in general complex i.e. $b_l=
|b_l|e^{i\phi_1}$, $e_l= |e_l|e^{i\phi_2}$. All the nonvanishing elements of $M_\nu$ are, in general, complex.
The charged lepton mass matrix $M_l$ can be diagonalized by the unitary transformation 
\begin{equation}
M_l=V_lM_l^dV_l^\dagger,
\end{equation}
where $V_l^\dagger=V_l^{-1}$. The hermitian matrix $M_l$ can, in
general, be written as
\begin{equation}
M_l=P_lM_l^{r}P_l^\dagger
\end{equation}
where $P_l$ is a unitary diagonal phase matrix,
$diag(1,e^{i\phi_1},e^{i\phi_2})$ and $M_l^{r}$ is a real matrix
which can be diagonalized by a real orthogonal matrix $O_l$ as
\begin{equation}
M_l^{r}=O_lM_l^{d}O_l^T
\end{equation}
where the superscript T denotes transposition and
$M_l^d$=$diag(m_e,m_{\mu}, m_{\tau})$. From Eqns. (10) and (11), the unitary
matrix $V_l$ is given by
\begin{equation}
V_l=P_lO_l.
\end{equation} 
Using the invariants, Tr$M_l^r$, Tr$M_l^{r^ 2}$ and Det$M_l^{r}$ we get
the matrix elements $f_l$, $|e_l|$ and $|b_l|$ as
\begin{eqnarray}
f_l=m_e - m_\mu + m_\tau -2a_l ,\nonumber\\
|e_l|=\left(-\frac{(2a_l-m_e+m_\mu)(2a_l-m_e-m_\tau)(2a_l+m_\mu -m_\tau)}{3a_l-m_e+m_\mu -m_\tau}\right)^\frac{1}{2},\\
|b_l|=\left(\frac{(a_l-m_e)(a_l+m _\mu)(a_l-m_\tau)}{3a_l-m_e+m_\mu -m_\tau} \right)^\frac{1}{2}.\nonumber
\end{eqnarray} 
Here, $a_l$ has two allowed ranges $\left(\frac{m_\tau-m_\mu}{2}\right) <a_l<\left(\frac{m_\tau+m_e}{2}\right)$ and $\left(-\frac{m_\mu}{2}<a_l<\frac{m_e}{2}\right)$ for the elements $|e_l|$ and $|b_l|$ to be real. 
The elements of the diagonalizing matrix, $O_l$ can be written in terms of the charged lepton masses $m_e$, $m_\mu$, $m_\tau$ and the charged lepton mass matrix elements $a_l$, $b_l$ and $e_l$. The elements $b_l$ and $c_l$ can be written in terms of $a_l$, thus, leading to a single unknown parameter $a_l$ and $O_l$ is given by
\begin{equation}
 O_l=\left(%
\begin{array}{ccc}
 \frac{(-e_l^2-(a_l-m_e)(2a_l+m_\mu- m_\tau))}{A} &  \frac{(e_l^2+(a_l+m_\mu)(2a_l-m_e- m_\tau))}{B}  & \frac{(e_l^2+(2a_l-m_e+ m_\mu)(a_l-m_\tau))}{C}\\ 
 \frac{b_l(2a_l+m_\mu- m_\tau)}{A}  & \frac{b_l(-2a_l+m_e+m_\tau)}{B}&\frac{b_l(-2a_l+m_e-m_\mu )}{C}\\
  \frac{b_l e_l}{A} &-\frac{b_le_l}{B} & -\frac{b_l e_l}{C} \\
\end{array}
\right)
\end{equation}
where A, B and C are given by
\begin{equation}
 \left.\begin{array}{c}
A=[4a_l^4 + b_l^2(e_l^2 + ( m_\mu - m_\tau)^2)- 4a_l^3(2m_e- m_\mu + m_ \tau)+(e_l^2-m_em_\mu+ m_em_\tau)^2\nonumber \\ + a_l^2 (4b_l^2+4e_l^2+4m_e^2+(m_\mu- m_\tau)^2+8m_e(-m_\mu + m_\tau))-2a_l(2b_l^2(-m_\mu +m_\tau)+ \nonumber\\
(2m_e-m_\mu+ m_\tau)(e_l^2+m_e(-m_\mu+ m_\tau)))]^\frac{1}{2}, \nonumber \\
B=[4a_l^4 - 4a_l^3(m_e - 2m_\mu + m_\tau)+ a_l^2(4b_l^2+4e_l^2-m_e^2-8m_em_\mu+ 4m_\mu^2+ 2m_em_\tau- \nonumber \\8m_\mu m_\tau+ m_\tau^2) + (e_l^2-m_\mu(m_e+m_\tau))^2+b_l^2(e_l^2+(m_e+m_\tau)^2)-2a_l(2b_l^2\nonumber\\
(m_e+m_\tau)+(m_e-2m_\mu+m_\tau)(e_l^2-m_\mu(m_e+m_\tau))]^\frac{1}{2},\nonumber\\
C=[4a_l^4 + b_l^2(e_l^2+(m_e-m_\mu)^2)- 4a_l^3(m_e-m_\mu+ 2m_\tau)+ (e_l^2+m_\tau(m_e-m_\tau))^2+\nonumber\\ a_l^2(4b_l^2+4e_l^2(m_e-m_\mu)^2+8(m_e-m_\mu)m_\tau+ 4m_\tau^2)-2a_l(2b_l^2(m_e-m_\mu) \nonumber\\
+(m_e-m_\mu+ 2m_\tau)(e_l^2+(m_e-m_\mu)m_\tau))]^\frac{1}{2}.
\end{array}\right\}
\end{equation}
If $a_l$ is known, the diagonalizing matrix $O_l$ and the real charged lepton mass matrix $M_l^r$ are fully determined since the charged lepton masses are known.
The complex symmetric neutrino mass matrix, $M_{\nu}$ is
diagonalized by a complex unitary matrix $V_{\nu}$:
\begin{equation}
M_\nu=V_\nu M_ \nu^{diag}V_\nu^T.
\end{equation}
The lepton mixing matrix or Pontecorvo-Maki-Nakagawa-Sakata
matrix, $U_{PMNS}$ \cite{23} is given by
\begin{equation}
U_{PMNS}=V_l^\dagger V_\nu .
\end{equation}
The mixing matrix $U_{PMNS}$ consists of three non
trivial $CP$ violating phases: the Dirac phase $\delta$ and the two
Majorana phases: $\alpha$, $\beta$ and the three neutrino mixing
angles viz. $\theta_{12}$, $\theta_{23}$ and $\theta_{13}$. The
neutrino mixing matrix can be written as the product of two matrices
characterizing Dirac-type and Majorana-type $CP$ violation \textit{i.e.}
\begin{equation}
U_{PMNS}=UP
\end{equation}
where $U$ and $P$ \cite{24} are given by
\begin{equation}
U=\left(
\begin{array}{ccc}
c_{12}c_{13} & s_{12}c_{13} & s_{13}e^{-i\delta } \\
-s_{12}c_{23}-c_{12}s_{23}s_{13}e^{i\delta } &
c_{12}c_{23}-s_{12}s_{23}s_{13}e^{i\delta } & s_{23}c_{13} \\
s_{12}s_{23}-c_{12}c_{23}s_{13}e^{i\delta } &
-c_{12}s_{23}-s_{12}c_{23}s_{13}e^{i\delta } & c_{23}c_{13}
\end{array}
\right),
 P = \left(
\begin{array}{ccc}
1 & 0 & 0 \\
0 & e^{i\alpha } & 0 \\
0 & 0 & e^{i\left( \beta +\delta \right) }
\end{array}
\right).
\end{equation}
From Eqns. (16) and (17), the neutrino mass matrix $M_\nu$ can be written as,
\begin{equation}
M_\nu=P_lO_lU_{PMNS}M_\nu^{diag}U_{PMNS}^TO_l^TP_l^\dagger.
\end{equation}
A texture zero and an equality in $M_\nu$ yield two complex equations
\begin{equation}
m_1ap+m_2bqe^{2i\alpha}+m_3cre^{2i(\beta+\delta)}=0,
\end{equation}
\begin{equation}
m_1(a^2-d^2)+m_2(b^2-g^2)e^{2i\alpha}+m_3(c^2-h^2)e^{2i(\beta+\delta)}=0,
\end{equation}
where the complex coefficients $a$, $b$, $c$, $d$, $g$, $h$, $p$, $q$ and $r$  are given by
\begin{eqnarray}
a=O_{11}U_{e1}+O_{12}U_{m1}+O_{13}U_{t1},\nonumber\\
b=O_{11}U_{e2}+O_{12}U_{m2}+O_{13}U_{t2},\nonumber\\
c=O_{11}U_{e3}+O_{12}U_{m3}+O_{13}U_{t3},\nonumber\\
d=O_{21}U_{e1}+O_{22}U_{m1}+O_{23}U_{t1},\nonumber\\
g=O_{21}U_{e2}+O_{22}U_{m2}+O_{23}U_{t2},\nonumber\\
h=O_{21}U_{e3}+O_{22}U_{m3}+O_{23}U_{t3},\nonumber\\
p=O_{31}U_{e1}+O_{32}U_{m1}+O_{33}U_{t1},\nonumber\\
q=O_{31}U_{e2}+O_{32}U_{m2}+O_{33}U_{t2},\nonumber\\
r=O_{31}U_{e3}+O_{32}U_{m3}+O_{33}U_{t3}.\nonumber\\
\end{eqnarray}

The two complex Eqns. (21) and (22) can be solved simultaneously to get the following two mass ratios
\begin{equation}
\begin{large}
\begin{array}{c}
\frac{m_1}{m_2}e^{-2i\alpha}=\left(\frac{cr(b^2-g^2)-bq(c^2-h^2)}{ap(c^2-h^2)-cr(a^2-d^2)}\right),\nonumber\\
\frac{m_1}{m_3}e^{-2i\beta}=\left(\frac{bq(c^2-h^2)-cr(b^2-g^2)}{ap(b^2-g^2)-bq(a^2-d^2)}\right)e^{2i\delta}.
\end{array}
\end{large}
\end{equation}
It is useful to enumerate the number of parameters in Eqn. (24). The nine
parameters including the three neutrino mixing angles
$(\theta_{12},\theta_{23},\theta_{13})$, three neutrino mass
eigenvalues $(m_1,m_2,m_3)$, two Majorana phases $(\alpha,\beta)$
and one Dirac-type $CP$ violating phase ($\delta$) come from the
neutrino sector and the four parameters including the three charged lepton masses
$(m_e,m_\mu,m_\tau)$ and $a_l$ come from the charged lepton sector,
thus, totalling 13 parameters. The three charged lepton masses are known \cite{25}
\begin{equation}
m_e=0.510998910 MeV, \\
m_{\mu}=105.658367 MeV, \\
m_{\tau}=1776.84 MeV. \\
\end{equation}
The masses
$m_{2}$ and $m_{3}$ can be calculated from the mass-squared
differences $\Delta m_{12}^{2}$ and $\Delta m_{23}^{2}$ using the
relations
\begin{equation}
m_{2}=\sqrt{m_{1}^{2}+\Delta m_{12}^{2}}
\end{equation}
and
\begin{equation}
m_{3}=\sqrt{m_{2}^{2}+\Delta m_{23}^{2}}.
\end{equation}
 Using the known values of the two mass-squared
 differences and the two mixing angles, we can constrain the other parameters.
The parameters $\delta$ and
$a_l$ are varied uniformly within their full possible ranges while $\theta_{13}$ is varied uniformly upto its upper bound given by CHOOZ \cite{20}. Thus, we are left with three unknown parameters viz. $m_1$, $\alpha$, $\beta$.
The magnitudes of the two mass ratios are given by
\begin{equation}
\begin{large}
\sigma=\left|\frac{m_1}{m_2}e^{-2i\alpha}\right|,\\
\end{large}
\end{equation}
and
\begin{equation}
\begin{large}
\rho=\left|\frac{m_1}{m_3}e^{-2i\beta}\right|.
\end{large}
\end{equation}
while the $CP$ violating Majorana phases are given by
\begin{equation}
\alpha=-\frac{1}{2}arg\left(\frac{cr(b^2-g^2)-bq(c^2-h^2)}{ap(c^2-h^2)-cr(a^2-d^2)}\right),
\end{equation}
\begin{equation}
\beta=-\frac{1}{2}arg\left(\frac{bq(c^2-h^2)-cr(b^2-g^2)}{ap(b^2-g^2)-bq(a^2-d^2)}\right)e^{2i\delta}.
\end{equation}
Since, $\Delta m_{12}^{2}$ and $\Delta m_{23}^{2}$ are known experimentally, the values of mass ratios $(\rho,\sigma)$ from Eqns. (28) and (29) can be used to calculate $m_1$.
This can be achieved by inverting Eqns. (26) and (27) to obtain the two values of $m_1$ viz.
\begin{equation}
m_{1}=\sigma \sqrt{\frac{ \Delta
m_{12}^{2}}{1-\sigma ^{2}}},
\end{equation}
and
\begin{equation}
m_{1}=\rho \sqrt{\frac{\Delta m_{12}^{2}+
\Delta m_{23}^{2}}{ 1-\rho^{2}}}.
\end{equation}
We vary the oscillation parameters within their known experimental ranges. The two values of $m_1$ obtained from the mass ratios $\rho$ and $\sigma$, respectively must be equal to within the errors of the oscillation parameters for this hybrid texture to be phenomenologically viable.\\ 
\begin{figure}
\begin{center}
\epsfig{file=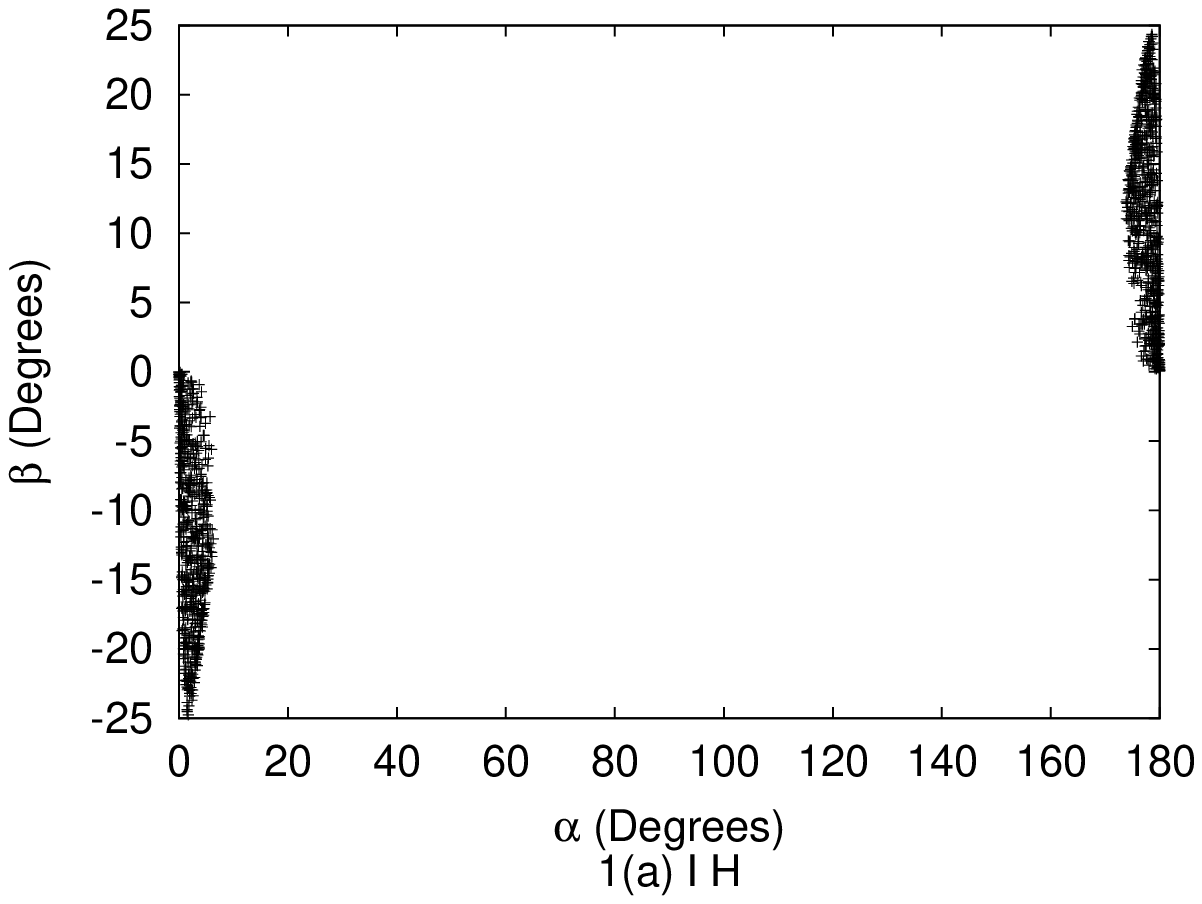,height=5.0cm,width=5.0cm}
\epsfig{file=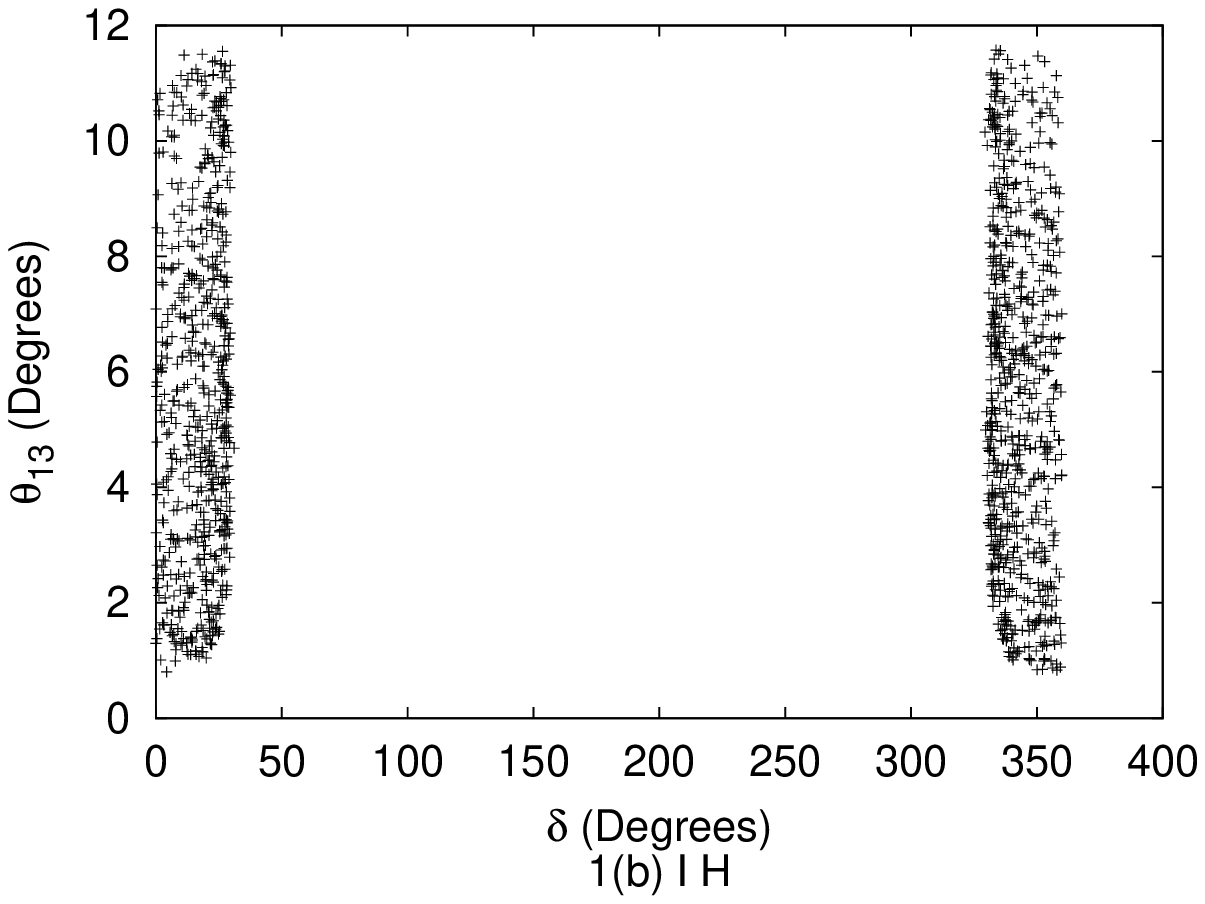,height=5.0cm,width=5.0cm}
\epsfig{file=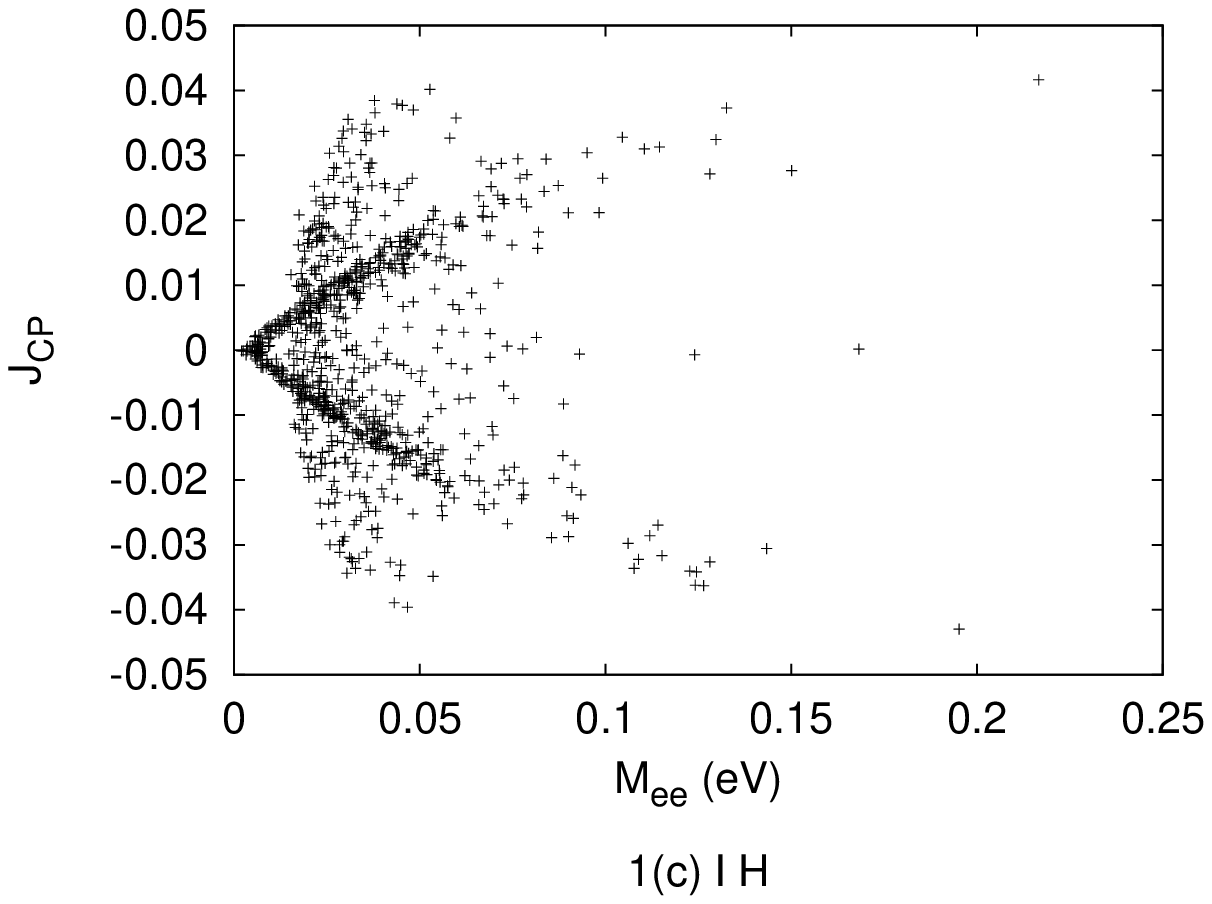,height=5.0cm,width=5.0cm}
\epsfig{file=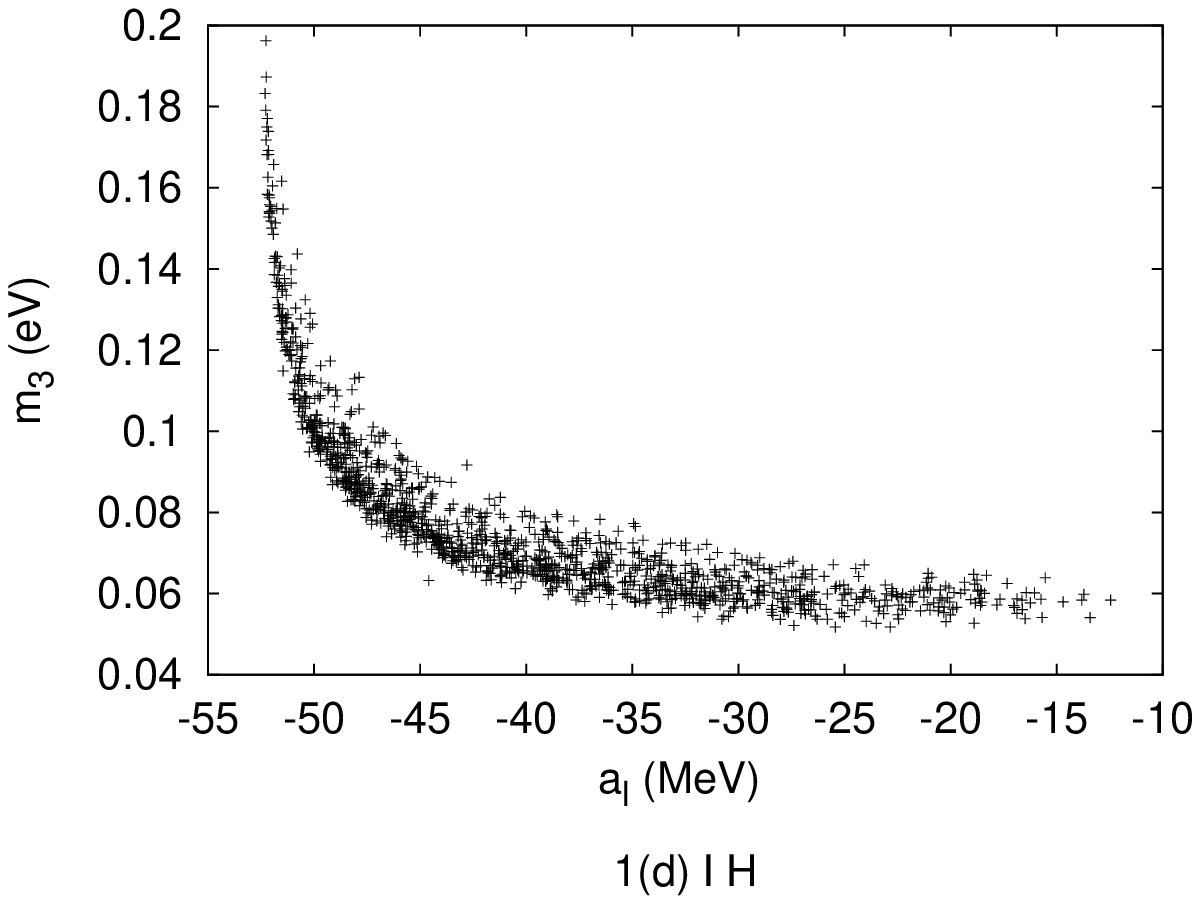,height=5.0cm,width=5.0cm}
\epsfig{file=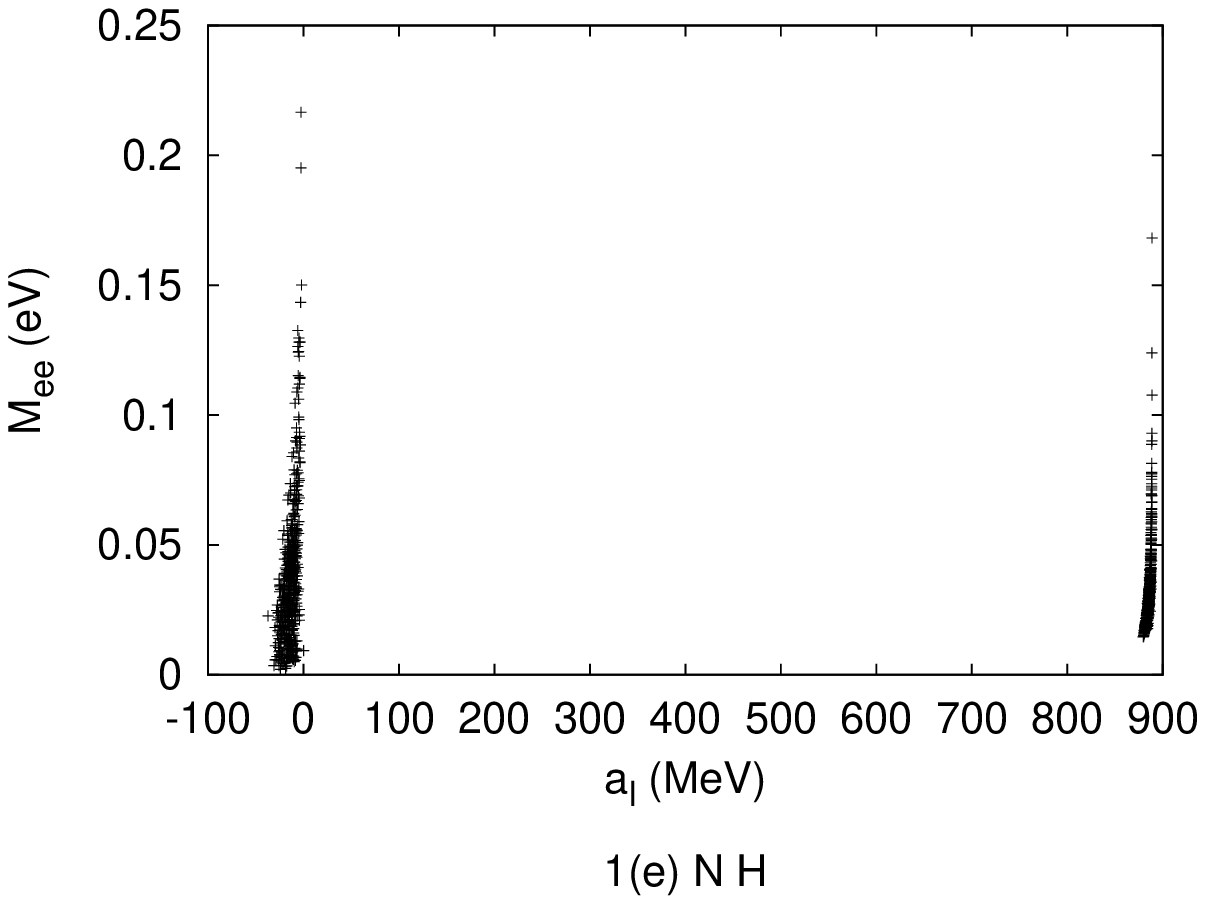,height=5.0cm,width=5.0cm}
\epsfig{file=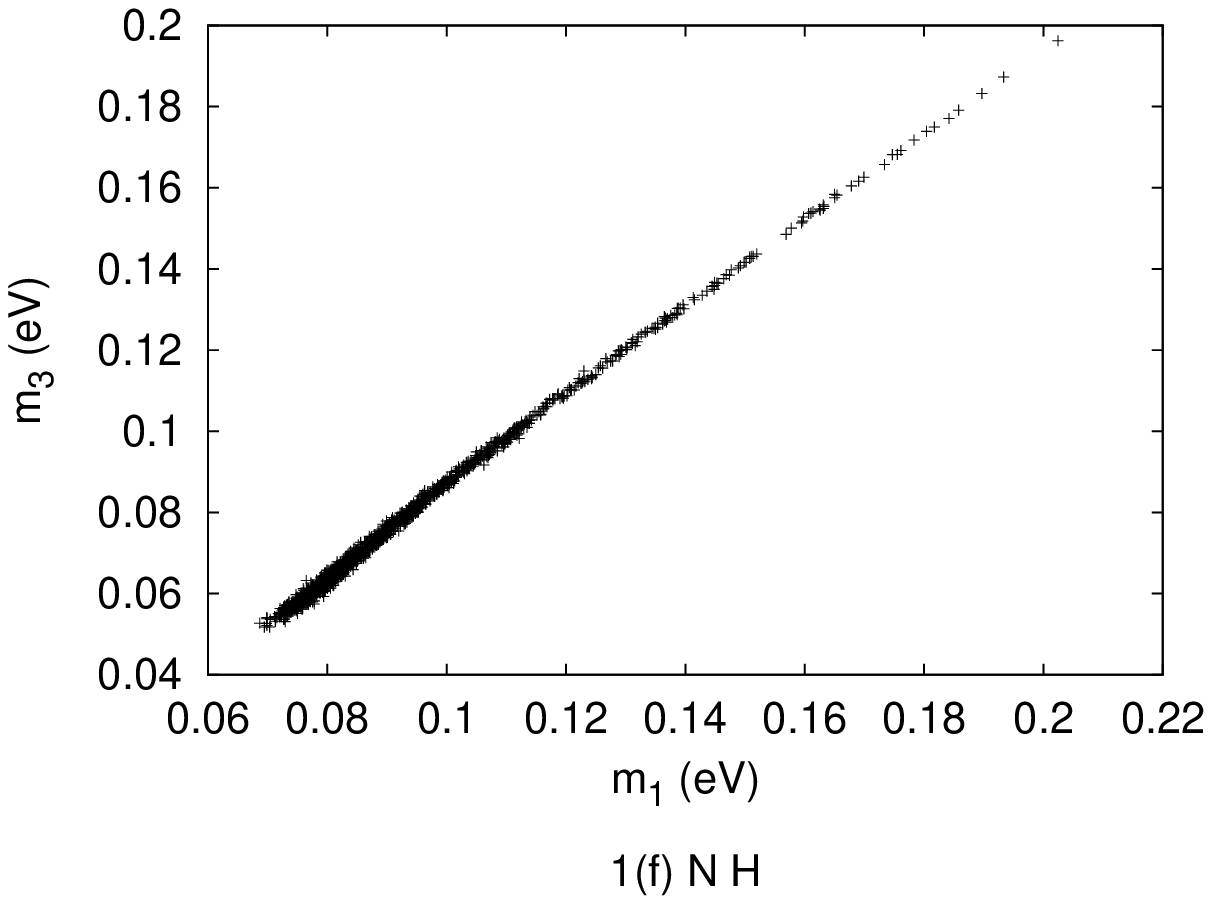,height=5.0cm,width=5.0cm}
\end{center}
\caption{Correlation plots for class I.}
\end{figure}
This class of hybrid texture has both normal and inverted hierarchical mass spectum. We get some interesting predictions for other parameters for each hierarchy which are to be probed in the forthcoming neutrino oscillation experiments. For inverted hierarchy, a small range  of the two Majorana-type $CP$ violating phases $\alpha$ and $\beta$ is allowed at 3$\sigma$. It can be seen from figure 1(a) that $\alpha$ can take the values $0^o$ or $180^o$ while $\beta$ is constrained to the range $(-25^o)-(25^o)$. There exists a clear bound on the reactor neutrino mixing angle ($\theta_{13}>1^o$) while the Dirac-type $CP$ violating phase $\delta$ is disallowed between $30^o - 130^o $ (Fig. 1(b)). There exists an upper bound of $0.2 eV$ on the effective Majorana mass $M_{ee}$ and the allowed range of Jarlskog rephasing invariant $J_{CP}$ is $(-0.04)-(0.04)$ (Fig. 1(c)). The range of the unknown parameter $a_l$ allowed by the current neutrino oscillation data is $[(-53)\leq a_l \leq (-10)]MeV$ as depicted in Fig. 1(d). For normal hierarchy an upper bound of $0.25 eV$ is obtained for the effective Majorana mass $M_{ee}$. Two highly constrained regions of parameter space are obtained for the free parameter $a_l$ as depicted in Fig. 1(e). All the other hybrid texture structures in this class have the same physical implications. 
\newpage
\subsection{Class II}
Here, we study the phenomenological implications of the parallel hybrid texture structure for the charged lepton mass matrix and the neutrino mass matrix with equality between (2,2) and (3,3) elements and a texture zero at (1,1) place (Case IIA): 
\begin{equation}
M_l=\left(%
\begin{array}{ccc}
  0 & b_l & c_l \\
  b_l^* & d_l & e_l \\
  c_l^* &e_l^* & d_l \\
\end{array}%
\right),
 M_\nu=\left(%
\begin{array}{ccc}
  0 & b_\nu & c_\nu \\
  b_\nu& d_\nu & e_\nu\\
  c_\nu & e_\nu  & d_\nu \\
\end{array}
\right).
\end{equation}
We consider the phases appearing in the charged lepton mass matrix to be factorizable since it is not possible to completely remove all the phases from this type of charged lepton mass matrix where both equality and zero appear along the diagonal entries. We perform analysis similar to class I for this hybrid texture.
Using the conditions from three invariants, we get the matrix elements $d_l$, $|b_l|$ and $|c_l|$ to be
\begin{eqnarray}
d_l=\frac{m_e-m_{\mu}+m_{\tau}}{2},\nonumber\\
|b_l|=\frac{(-4e_l^2(m_e-m_\mu + m_\tau)+(m_e-m_\mu - m_\tau)(m_e+m_\mu - m_\tau)(m_e+m_\mu +m_\tau))}{e_lX},\\
|c_l|=\frac{1}{X}.\nonumber
\end{eqnarray} 
where 
\begin{eqnarray}
X= 4[ (-8e_l^2+\frac{1}{e_l}((2e_l-m_e-m_\mu -m_\tau)(2e_l+m_e-m_\mu -m_\tau)\nonumber \\ (2e_l+m_e+m_\mu -m_\tau)(2e_l-m_e-m_\mu +m_\tau)(2e_l-m_e+m_\mu +m_\tau)\\(2e_l+m_e+m_\mu +m_\tau))^\frac{1}{2})+2(m_e^2+2m_e(m_\mu- m_\tau)+(m_\mu+m_\tau)^2)]^\frac{1}{2} \nonumber .
\end{eqnarray}
Here, $e_l$ should be in the range  $\frac{(m_\tau-m_\mu-m_e)}{2} <e_l< \frac{(m_\tau+m_\mu-m_e)}{2}$ for the elements $b_l$ and $c_l$ to be real.
The elements of the diagonalizing matrix $O_l$ can be written in terms of the charged lepton masses and the parameters $e_l$, $b_l$ and $c_l$. The parameters $b_l$ and $c_l$ are the functions of $e_l$ (Eqn. (35)), thus, leading to a single unknown parameter $e_l$ and $O_l$ is given by
\begin{equation}
\begin{large}
 O_l=\left(%
\begin{array}{ccc}
  -\frac{(2b_le_l+c_l(m_e+m_\mu - m_\tau))}{A} & \frac{(2b_le_l-c_l(m_e+m_\mu + m_\tau))}{B}  &\frac{(2b_le_l+c_l(-m_e+m_\mu + m_\tau)))}{D}\\ 
 -\frac{2(b_lc_l+e_lm_e)}{A}  &\frac{2(b_lc_l-e_lm_\mu)}{B}&\frac{2(b_lc_l+e_lm_\tau)}{D}\\
  \frac{(2b_l^2-m_e(m_e+ m_\mu- m_ \tau))}{A} &\frac{(-2b_l^2+ m_\mu(m_e+ m_\mu+ m_ \tau))}{B} & \frac{(-2b_l^2+m_\tau (-m_e+ m_\mu + m_ \tau))}{D} \\
\end{array}
\right)
\end{large}
\end{equation}
where A, B, D are given by
\begin{equation}
 \left.\begin{array}{c}
A=[4b_l^4 + 4b_l^2(c_l^2 + e_l^2 - m_e(m_e + m_\mu - m_\tau))+ m_e^2 (4e_l^2 + (m_e+ m_\mu - m_ \tau)^2)\nonumber \\ + c_l^2 (m_e+ m_\mu - m_ \tau)^2 + 4b_lc_le_l(3m_e+m_\mu - m_\tau)]^\frac{1}{2}, \nonumber \\
B=[4b_l^4 + c_l^2(m_e + m_\mu + m_\tau)^2- 4b_lc_le_l(m_e+ 3m_\mu + m_ \tau) + 4b_l^2 (c_l^2+e_l^2-m_\mu)\nonumber \\(m_e+ m_\mu + m_ \tau) + m_\mu^2(4e_l^2 + (m_e + m_\mu + m_\tau))^2]^\frac{1}{2},\nonumber\\
D=[4b_l^4 + c_l^2(-m_e + m_\mu + m_\tau)^2 + 4b_lc_le_l(-m_e+ m_\mu +3m_ \tau)+ 4b_l^2 (c_l^2 + e_l^2\nonumber \\  + m_\tau(m_e - m_ \mu - m_\tau)+ m_\tau^2(4e_l^2+(-m_e + m_\mu + m_\tau))^2]^\frac{1}{2} 
\end{array}\right\}.
\end{equation}
This structure of hybrid texture of $M_\nu$ results in two complex equations
\begin{equation}
m_1a^2+m_2b^2e^{2i\alpha}+m_3c^2e^{2i(\beta+\delta)}=0,
\end{equation}
\begin{equation}
m_1(d^2-p^2)+m_2(g^2-q^2)e^{2i\alpha}+m_3(h^2-r^2)e^{2i(\beta+\delta)}=0.
\end{equation}
where the complex coefficients $a$, $b$, $c$, $d$, $g$, $h$, $p$, $q$ and $r$ have the same form as given in Eqn. (23).
The mass ratios can be found from the two complex Eqns. (39) and (40) and are given by
\begin{equation}
\begin{large}
\begin{array}{c}
\frac{m_1}{m_2}e^{-2i\alpha}=\left(\frac{c^2(g^2-q^2)-b^2(h^2-r^2)}{a^2(h^2-r^2)-c^2(d^2-p^2)}\right),\\
\frac{m_1}{m_3}e^{-2i\beta}=\left(\frac{b^2(h^2-r^2)-c^2(g^2-q^2)}{a^2(g^2-q^2)-b^2(d^2-p^2)}\right)e^{2i\delta}.
\end{array}
\end{large}
\end{equation}
The absolute values of Eqn. (41) yield the two mass ratios $(\frac{m_1}{m_2})$ and $(\frac{m_1}{m_3})$ while the arguments of these equations give us information about the two Majorana type CP violating phases $\alpha$ and $\beta$ as shown in detail earlier.
By equating the two values of $m_1$ to within the errors of the oscillation parameters we obtain interesting implications for this hybrid texture. \\
Specifically, both normal and hierarchies are allowed for this class. For inverted hierarchy, the Majorana-type CP violating phase $\alpha$ is constrained to the range:$(75^o-105^o)$(Fig. 2(a)). In this case, an upper as well as a lower bound is obtained for effective Majorana mass, $(0.01<M_{ee}<0.08)eV$ and a highly constrained range $(880-940)MeV$ for the free parameter $e_l$ is allowed as can be seen from figure 2(b). For normal hierarhy $M_{ee}$ is constrained to be less than $0.1 eV$ (Fig. (2c)). 

\begin{figure}
\begin{center}
\epsfig{file=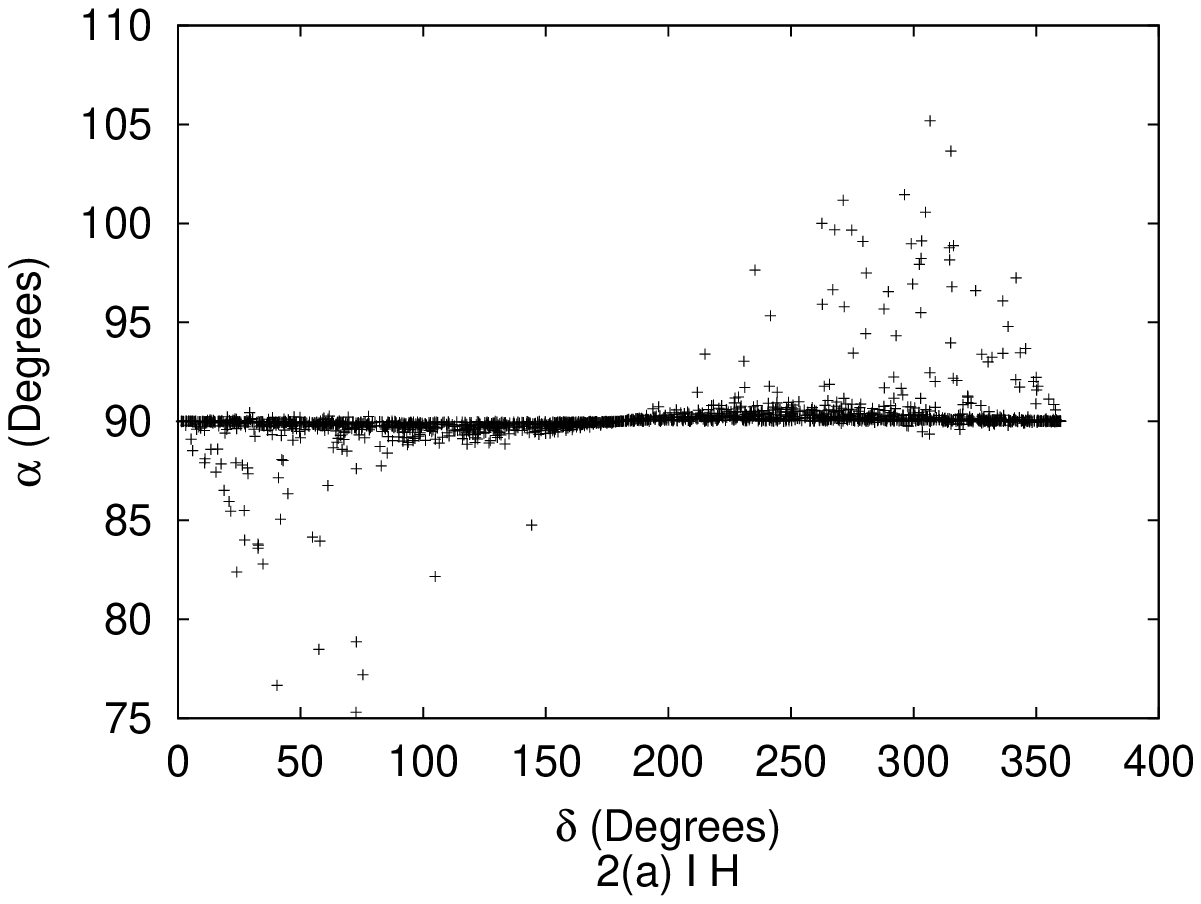,height=5.0cm,width=5.0cm}
\epsfig{file=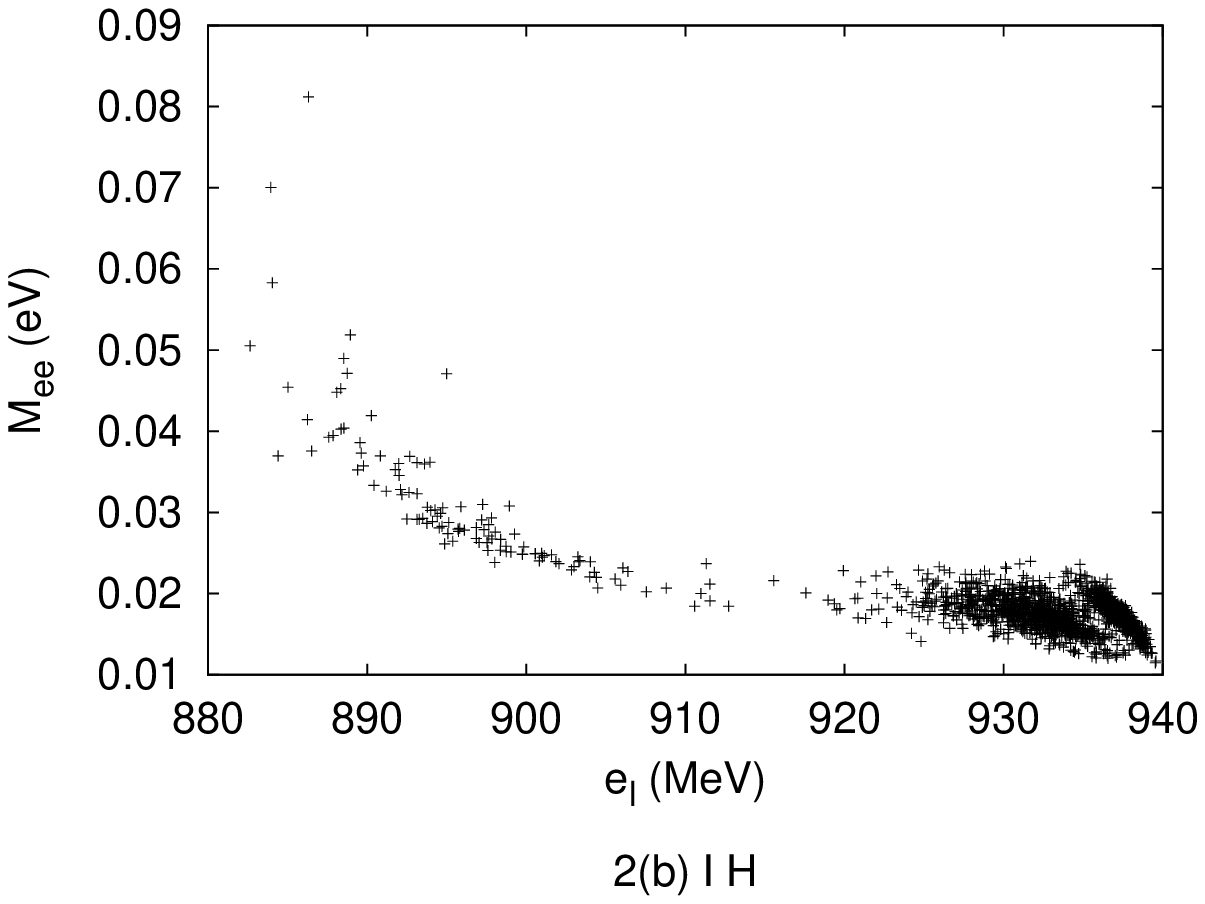,height=5.0cm,width=5.0cm}
\epsfig{file=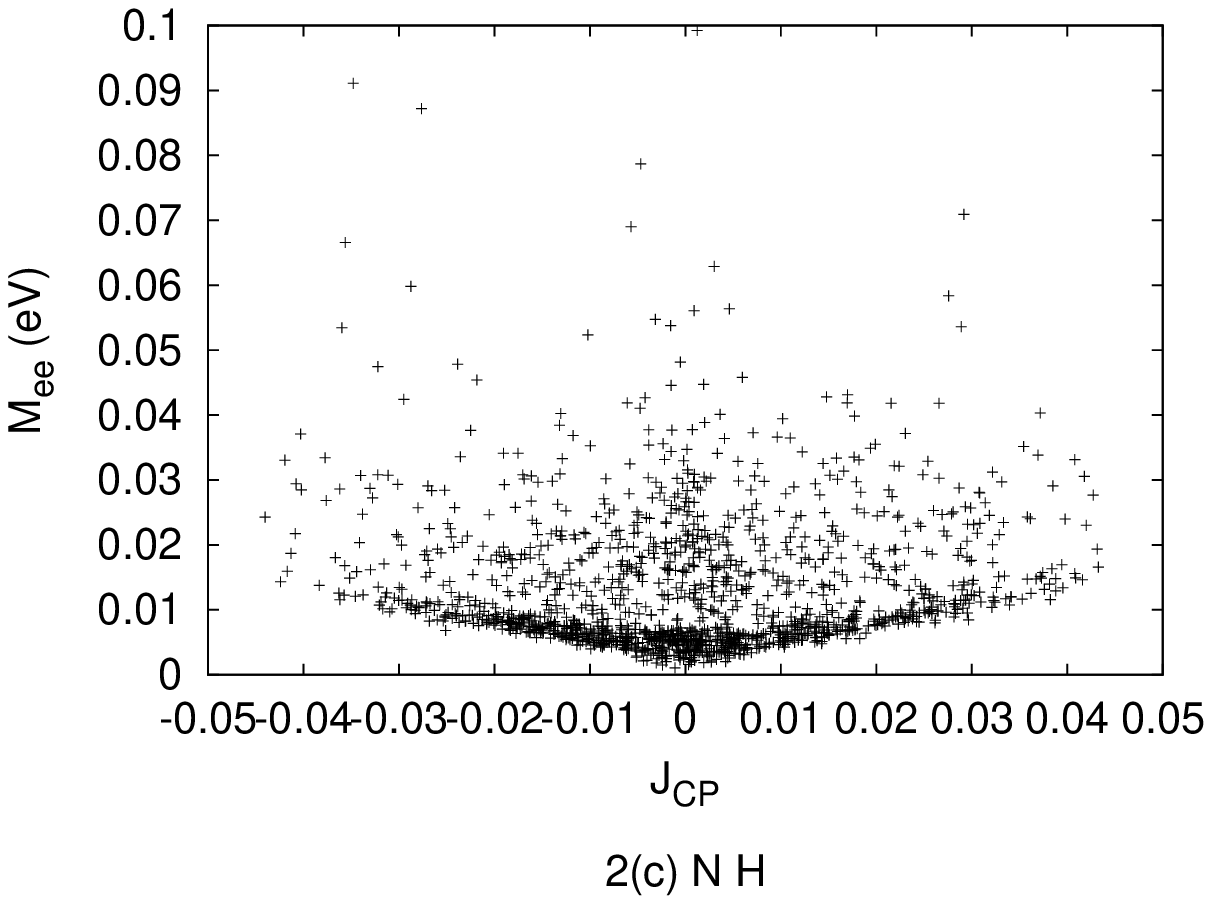,height=5.0cm,width=5.0cm}
\end{center}
\caption{Correlation plots for class II.}
\end{figure}
\subsection{Class III}
Another class of hybrid textures which leads to interesting implications is when the texture zero is at (1,1) position while (1,2) and (1,3) elements are equal (Case IIIA). Here, 
\begin{equation}
M_l=\left(%
\begin{array}{ccc}
  0 & b_l & b_l \\
  b_l^* & d_l & e_l \\
  b_l^* &e_l^* & f_l \\
\end{array}%
\right),
 M_\nu=\left(%
\begin{array}{ccc}
  0 & b_\nu & b_\nu \\
  b_\nu& d_\nu & e_\nu\\
 b_\nu & e_\nu  & f_\nu \\
\end{array}
\right).
\end{equation}
Using the invariants, we get
the matrix elements $|b_l|$, $d_l$ and $f_l$ as  
\begin{eqnarray}
|b_l|=\sqrt{\frac{m_e m_\mu m_\tau}{(-2e_l+m_e-m_\mu + m_\tau)}},\nonumber\\
d_l=\frac{1}{2}\left[(\frac{(2e_l-m_e-m_\mu- m_\tau)(2e_l+m_e+m_\mu - m_\tau)(2e_l-m_e+m_\mu + m_\tau)}{(-2e_l+m_e-m_\mu +m_\tau)}\right]^\frac{1}{2}
 \nonumber\\+\frac{1}{2}(m_e-m_\mu + m_\tau),\nonumber\\
f_l= m_e-m_{\mu}+m_{\tau}-d_l.\nonumber\\
\end{eqnarray}
The free parameter $e_l$ is constrained to the range $\left(\frac{(m_e-m_\mu-m_\tau)}{2}\right) <e_l<\left(\frac{(m_\tau-m_\mu-m_e)}{2}\right)$ for the element $b_l$ to be real. The orthogonal diagonalizing matrix $O_l$ can be written in terms of the charged lepton masses and charged lepton mass matrix elements:
\begin{equation}
\begin{large}
O_l=\left(%
\begin{array}{ccc}
  -\frac{(e_l^2+(d_l-m_e)(d_l+m_\mu - m_\tau))}{A} & \frac{e_l^2+(d_l+m_\mu)(d_l-m_e+m_\tau)}{B}  &\frac{-(e_l^2+(d_l-m_e+m_\mu)(d_l-m_\tau))}{D}\\ 
\frac{b_l(d_l+e_l+m_\mu - m_\tau)}{A}  &\frac{b_l(-d_l-e_l+m_e+m_\tau)}{B}&\frac{b_l(d_l+e_l-m_e+m_\mu)}{D}\\
  \frac{b_l(-d_l+e_l+m_e)}{A} &\frac{b_l(d_l-m_e+m_\mu)}{B} & \frac{b_l(-d_l+e_l+m_\tau)}{D} \\
\end{array}
\right)
\end{large}
\end{equation}
where $A$, $B$ and $D$ are given by
\begin{equation}
 \left.\begin{array}{c}
A=[(e_l^2+(d_l-m_e)(d_l+m_\mu - m_\tau))^2+b_l^2(2d_l^2+2e_l^2+m_e^2+(m_\mu -m_\tau)^2+\\2e_l(m_e+m_\mu -m_\tau)-2d_l(m_e-m_\mu + m_\tau))], \nonumber \\
B=[(e_l^2+(d_l+m_\mu)(d_l-m_e-m_\tau))^2+b_l^2(2d_l^2+2e_l^2+m_\mu^2 +(m_e+m_\tau)^2-\\2d_l(m_e-m_\mu +m_\tau)-2e_l(m_e+m_\mu - m_\tau))]^\frac{1}{2},\nonumber\\
D=[(e_l^2+(d_l-m_e+m_\mu)(d_l-m_\tau))^2+b_l^2(2d_l^2+2e_l^2+m_\tau^2 +(m_e-m_\mu)^2-\\2d_l(m_e-m_\mu +m_\tau)+2e_l(-m_e+m_\mu + m_\tau))]^\frac{1}{2},\nonumber\\ 
\end{array}\right\}.
\end{equation}
The simultaneous existence of a texture zero and an equality in $M_\nu$ leads to the following complex equations
\begin{equation}
m_1a^2+m_2b^2e^{2i\alpha}+m_3c^2e^{2i(\beta+\delta)}=0,
\end{equation}
\begin{equation}
m_1(ad-ap)+m_2(bg-bq)e^{2i\alpha}+m_3(ch-cr)e^{2i(\beta+\delta)}=0.
\end{equation}
The complex coefficients are given in Eqn. (23). Using these two complex equations we find the two mass ratios to be
\begin{equation}
\begin{large}
\begin{array}{c}
\frac{m_1}{m_2}e^{-2i\alpha}=\left(\frac{c^2(bg-bq)-b^2(ch-cr)}{a^2(ch-cr)-c^2(ad-ap)}\right),\\
\frac{m_1}{m_3}e^{-2i\beta}=\left(\frac{b^2(ch-cr)-c^2(bg-bq)}{a^2(bg-bq)-b^2(ad-ap)}\right)e^{2i\delta}.
\end{array}
\end{large}
\end{equation}
We perform a similar numerical analysis for this class and find that it is consistent with normal hierarchy only. A stringent bound on $\theta_{13}$ is obtained. There exists a lower bound of $4^o$ on the 1-3 mixing angle (Fig.3(a)). The unknown parameter $e_l$ has two allowed regions viz. $[(-960)-(-680)]MeV$ and $(430-830)MeV$. The effective Majorana mass is constrained to be less than $0.0045 eV$ for both regions of $e_l$ (Fig. 3(b)).
\begin{figure}
\begin{center}  
\epsfig{file=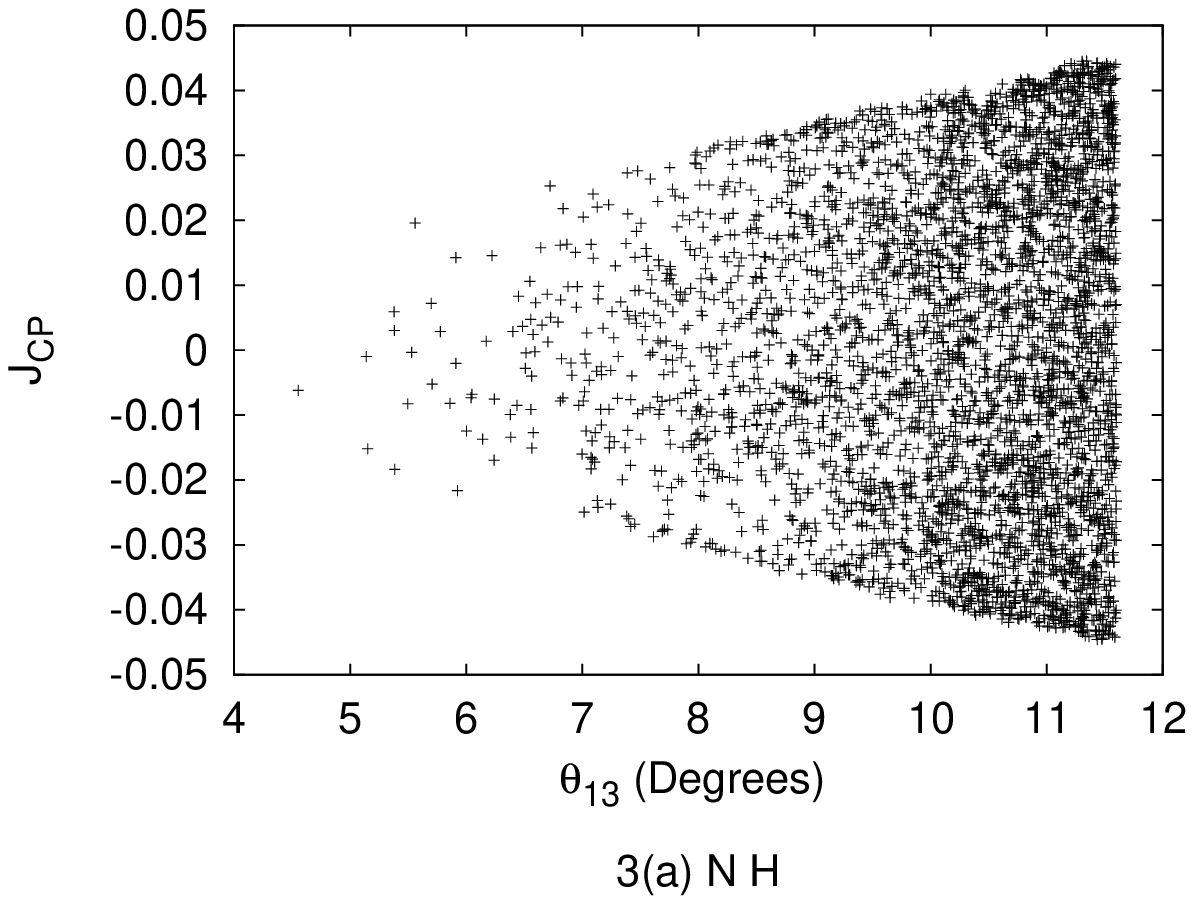,height=6.0cm,width=6.0cm}
\epsfig{file=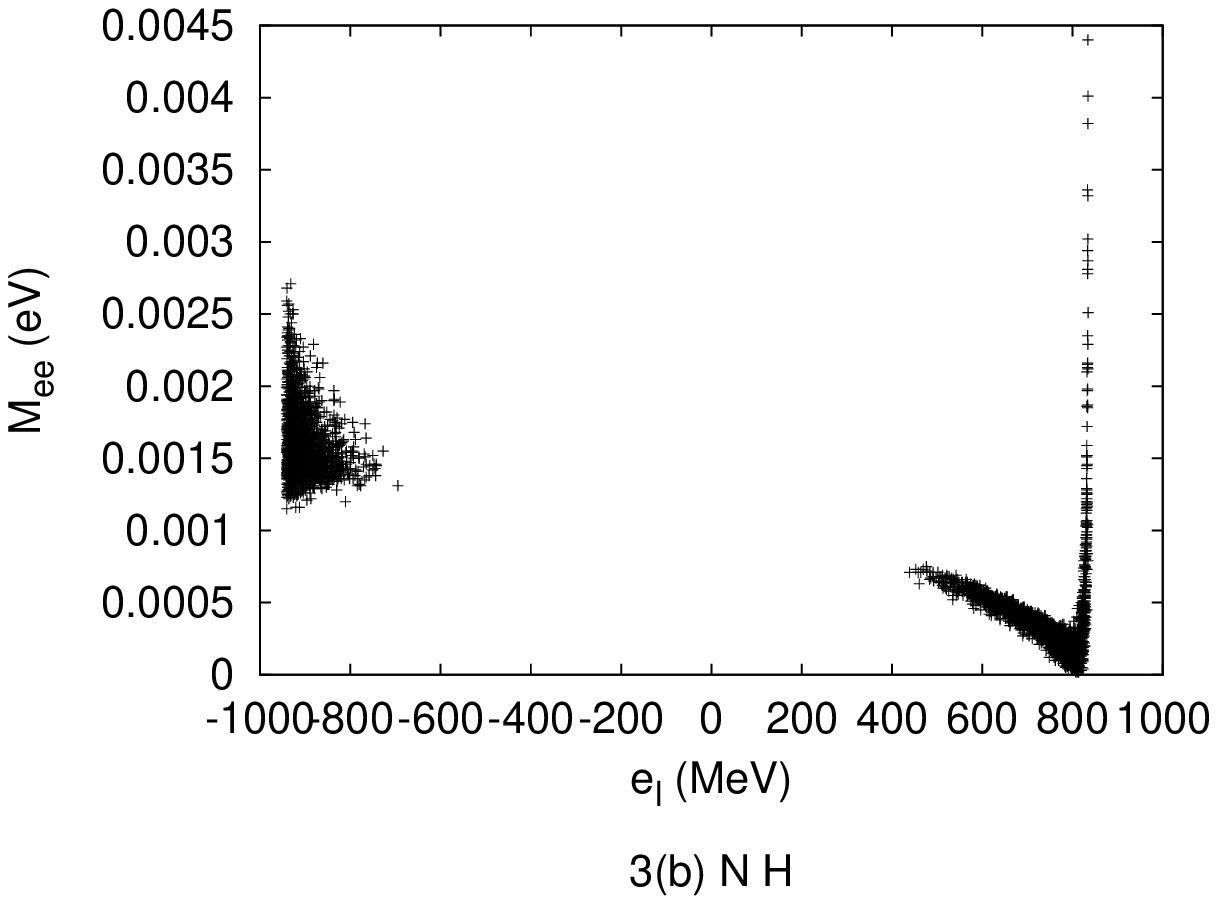,height=6.0cm,width=6.0cm}
\end{center}
\caption{Correlation plots for class III.}
\end{figure}
\section{Remaining viable classes of hybrid textures.}
The remaining phenomenologically viable classes are  IV, V, VI and VII. As for the classes discussed so far, the condition of a texture zero and an equality in $M_\nu$ results in two complex equations given in Table 2. The two mass ratios obtained for these hybrid textures are given in Table 3. When we apply our numerical analysis on all these classes we find that they have normal hierarchical mass spectra .\\ 
\begin{table}
\begin{large}
\begin{center}
\begin{tabular}{|c|c|c|}
\hline Case &Complex equations\\ 
\hline IVA & $m_1d^2+m_2g^2e^{2i\alpha}+m_3h^2e^{2i(\beta+\delta)}=0$\\ &$m_1(a^2-ad)+m_2(b^2-bg)e^{2i\alpha}+m_3(c^2-ch)e^{2i(\beta+\delta)}=0$ \\ 
\hline VA &$m_1p^2+m_2q^2e^{2i\alpha}+m_3r^2e^{2i(\beta+\delta)}=0$\\&$m_1(a^2-ad)+m_2(b^2-bg)e^{2i\alpha}+m_3(c^2-ch)e^{2i(\beta+\delta)}=0$ \\ 
\hline VIA & $m_1d^2+m_2g^2e^{2i\alpha}+m_3h^2e^{2i(\beta+\delta)}=0$\\& $m_1(a^2-dp)+m_2(b^2-gq)e^{2i\alpha}+m_3(c^2-hr)e^{2i(\beta+\delta)}=0$\\ 
\hline VIIA & $m_1ap+m_2bqe^{2i\alpha}+m_3cre^{2i(\beta+\delta)}=0$\\ & $m_1(a^2-ad)+m_2(p^2-pg)e^{2i\alpha}+m_3(b^2-bh)e^{2i(\beta+\delta)}=0$ \\ 
\hline 
\end{tabular}
\caption{Complex equations for remaining viable cases.}
\end{center}
\end{large}
\end{table} 
For class IV a constrained region $[(-55)-(165)] MeV$ for the free parameter $e_l$ along with an upper bound of $0.035 eV$ on effective Majorana mass $M_{ee}$ is obtained which can be seen from Fig. 4(a). For class V, two regions of solutions $(-100-0)MeV$ and $(833-930)MeV$ are obtained for the free parameter $a_l$. There exists an upper bound of $0.012 eV$on the effective Majorana mass $(M_{ee}$ (Fig.4(b)). There is a strong correlation between the two phases $\beta$ and $\delta$ as can be seen from Fig. 4(c).    
For class VI a stringent bound on effective Majorana mass is obtained: $(0.008<M_{ee}<0.04)eV$ Fig. 4(d).
Class VII is only marginally allowed since only 10-15 points are allowed whereas the total number of points generated in our numerical analysis is $10^7$.  

\begin{table}
\begin{large}
\begin{center}
\begin{tabular}{|c|c|c|}
\hline Case & $\frac{m_1}{m_3}e^{-2i\beta}$ & $\frac{m_1}{m_2}e^{-2i\alpha}$ \\ 
\hline IVA & $\left(\frac{g^2(c^2-ch)-h^2(b^2-bg)}{d^2(b^2-bg)-g^2(a^2-ad)}\right)$ &$\left(\frac{h^2(b^2-bg)-g^2(a^2-ad)}{d^2(c^2-ch)-h^2(a^2-ad)}\right)e^{2i\delta}$ \\ 
\hline VA & $\left(\frac{q^2(c^2-ch)-r^2(b^2-bg)}{p^2(b^2-bg)-q^2(a^2-ad)}\right)$ & $\left(\frac{r^2(b^2-bg)-q^2(c^2-ch)}{p^2(c^2-ch)-r^2(a^2-ad)}\right)e^{2i\delta}$ \\ 
\hline VIA & $\left(\frac{g^2(c^2-hr)-h^2(b^2-gq)}{d^2(b^2-gq)-g^2(a^2-dp)}\right)$ & $\left(\frac{h^2(b^2-gq)-g^2(c^2-hr)}{d^2(c^2-hr)-h^2(a^2-dp)}\right)e^{2i\delta}$ \\ 
\hline VIIA & $\left(\frac{bq(hb-b^2)-cr(gp-p^2)}{ap(gp-p^2)-bq(da-a^2)}\right)$ & $\left(\frac{cr(gp-p^2)-bq(hb-b^2)}{ap(hb-b^2)-cr(da-a^2)}\right)e^{2i\delta}$ \\ 
\hline 
\end{tabular}
\caption{Mass ratios for remaining viable cases.}
\end{center}
\end{large} 
\end{table}
\begin{figure}
\begin{center}  
\epsfig{file=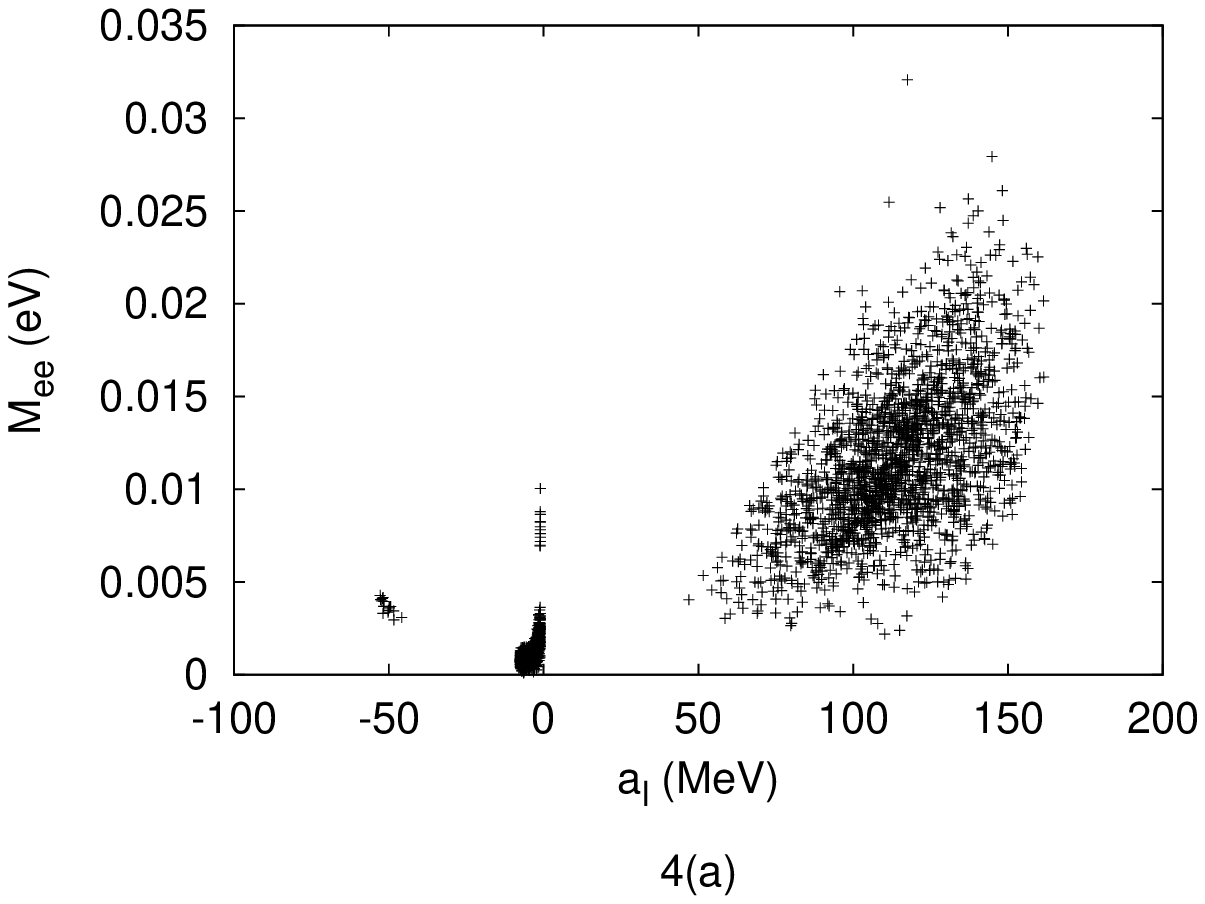,height=5.0cm,width=5.0cm}
\epsfig{file=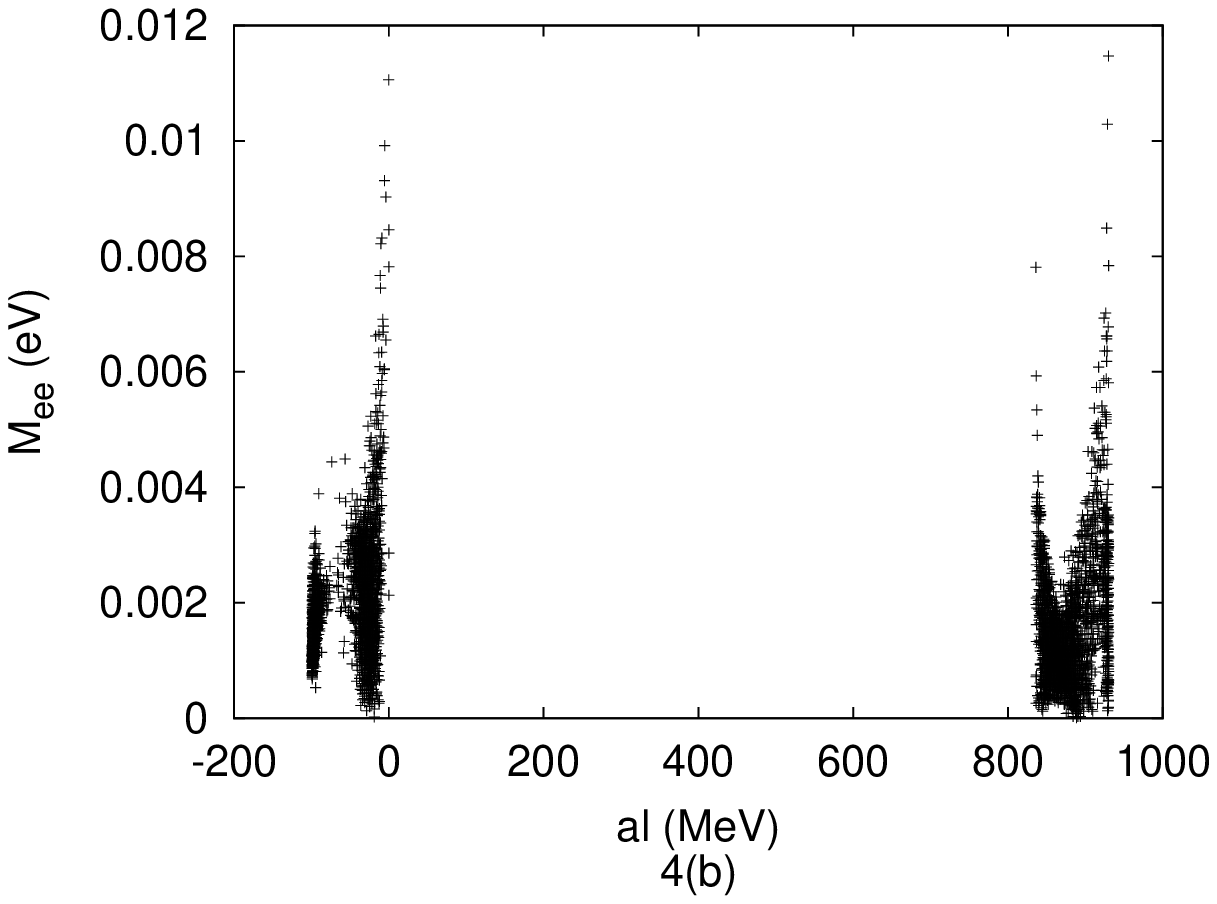,height=5.0cm,width=5.0cm}
\epsfig{file=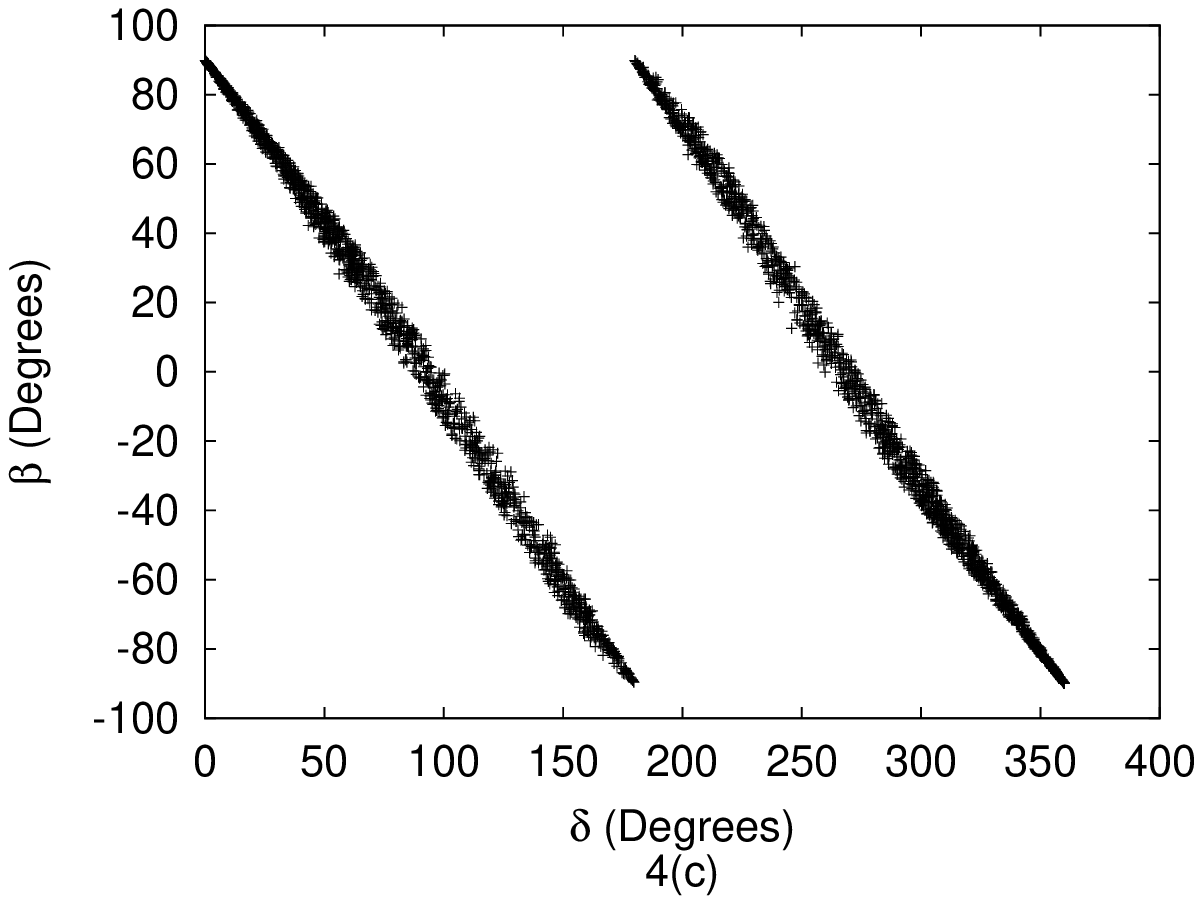,height=5.0cm,width=5.0cm}
\epsfig{file=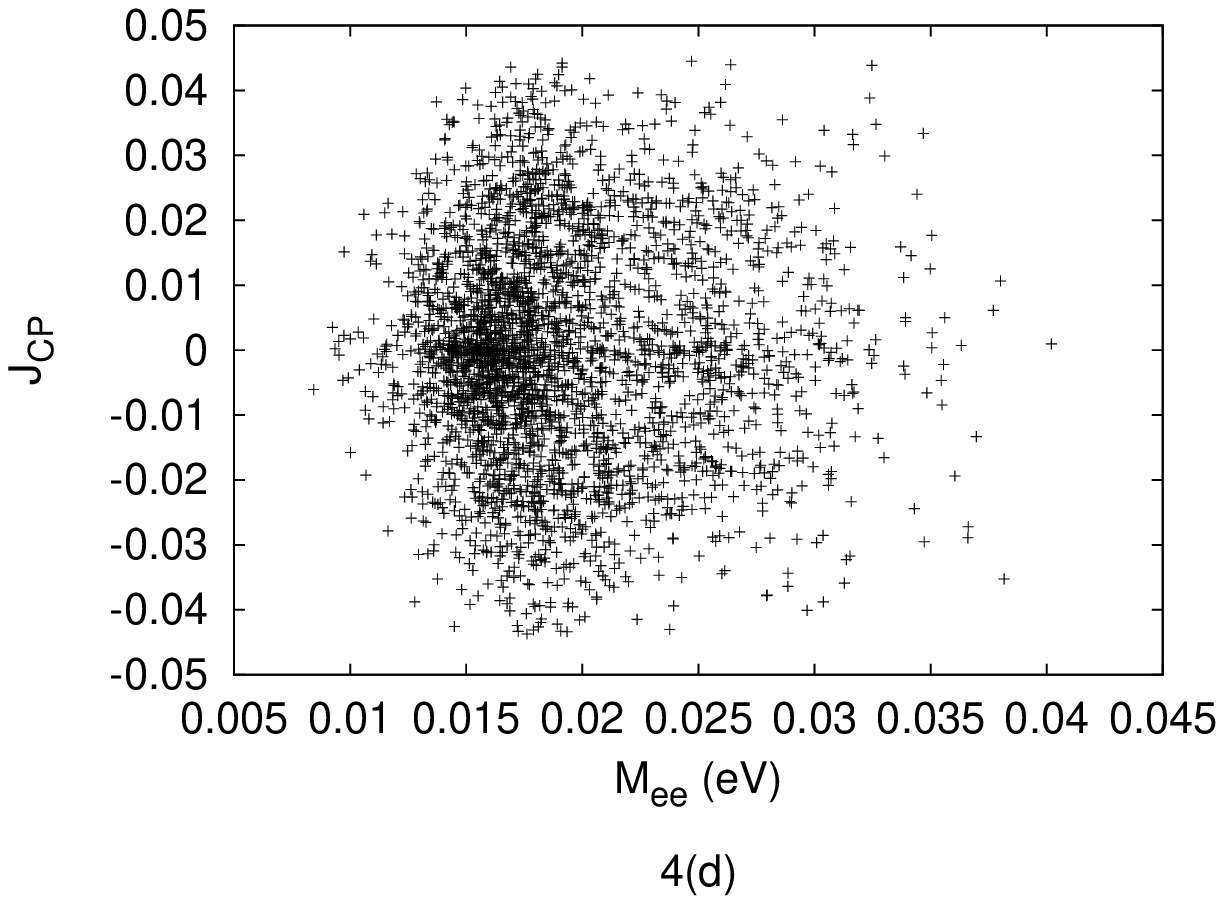,height=5.0cm,width=5.0cm}
\end{center}
\caption{Correlation plots for class IV, V and VI.}
\end{figure}

\section{Conclusions}
We presented a comprehensive phenomenological analysis for the parallel hybrid texture structures of the charged lepton and the neutrino mass matrices. These parallel hybrid texture structures cannot be obtained from arbitrary hermitian charged lepton and complex symmetric neutrino mass matrices through weak basis transformations and thus have physical implications. All the possible sixty hybrid texture structures are grouped into twelve classes using the permutation matrices. All textures in a class have the same physical implications. Six out of total twelve classes are found to be phenomenologically viable and have interesting implications while class VII is only marginally allowed. The remaining five classes of hybrid textures are phenomenologically disallowed. For each class, we obtained the allowed range for the only free parameter from the charged lepton sector. Predictions for the 1-3 mixing angle, the Dirac-type and Majorana-type CP violating phases are obtained for some of the allowed hybrid texture structures. We, also, obtained bounds on the effective Majorana mass  for all the allowed hybrid texture structures. The study of these parameters is significant as they are expected to be probed in the forthcoming experiments. 

\textbf{\textit{\Large{Acknowledgements}}}

The research work of S. D. is supported by the University Grants
Commission, Government of India \textit{vide} Grant No. 34-32/2008
(SR).  S. G. and R. R. G. acknowledge the financial support provided by the Council for Scientific and Industrial Research (CSIR), Government of India.

\end{document}